# Guidelines for the design of evolve and resequencing studies

Robert Kofler[1] and Christian Schlötterer[1]*

(1) Institut für Populationsgenetik, Vetmeduni Vienna, Veterinärplatz 1, 1210 Wien, Austria

* corresponding author

7/18/13

**Abstract**

Standing genetic variation provides a rich reservoir of potentially useful mutations facilitating the adaptation to novel environments. Experimental evolution studies have demonstrated that rapid and strong phenotypic responses to selection can also be obtained in the laboratory. When combined with the Next Generation Sequencing technology, these experiments promise to identify the individual loci contributing to adaption. Nevertheless, until now, very little is known about the design of such evolve and resequencing (E&R) studies. Here, we use forward simulations of entire genomes to evaluate different experimental designs that aim to maximize the power to detect selected variants. We show that low linkage disequilibrium in the starting population, population size, duration of the experiment and the number of replicates are the key factors in determining the power and accuracy of E&R studies. Furthermore, replication of E&R is more important for detecting the targets of selection than increasing the population size. Using an optimized design beneficial loci with a selective advantage as low as $s=0.005$ can be identified at the nucleotide level. Even when a large number of loci are selected simultaneously, up to 56% can be reliably detected without incurring large numbers of false positives. Our computer simulations suggest that, with an adequate experimental design, E&R studies are a powerful tool to identify adaptive mutations from standing genetic variation and thereby provide an excellent means to analyze the trajectories of selected alleles in evolving populations.

## Introduction

The importance of standing genetic variation to the adaptation of natural populations is well recognized [1-3]. Experimental evolution studies have successfully demonstrated that standing genetic variation enables a rapid phenotypic response in laboratory populations[4-6]. The advent of Next Generation Sequencing (NGS) has generated renewed interest in experimental evolution as it now has become possible to identify the loci contributing to adaptation [4, 5, 7, 8]. This approach, which has been termed evolve & resequence (E&R [6]), promises to unify two branches of genetics that have been separated for most of the 20th century: molecular genetics and population genetics [9]. E&R studies have the potential to identify the loci contributing to adaptation (molecular genetics) as well as to analyze the trajectories of these loci during adaptation (population genetics). At present little is known on how the design of E&R studies can be optimized to identify the maximal number of selectively favored loci [10]. In fact it is not yet known whether a few causal loci can be distinguished from millions of neutrally evolving ones, raising the important question whether E&R studies can work at all. Here, we address these questions by forward simulations of entire genomes and evaluate the power of different experimental designs to identify beneficial loci. We base the simulations on *Drosophila melanogaster*, since it is widely used in experimental evolution studies [4, 5, 11-13]. We demonstrate that the experimental design has a pronounced influence on the power to detect selected loci. While the major challenge for weakly selected loci is the distinction from unlinked neutrally evolving ones, strongly selected loci frequently result in a large number of linked false positives. Increasing the number of replicates, population size and the duration of the experiment increases the power of E&R studies. Using such an optimized design up to 56% of the selected loci can be detected at the resolution of a single nucleotide. Even weakly selected loci with a selective advantage as low as $s=0.005$ can be identified in the laboratory. Our computer simulations suggest that, provided an adequate experimental design, E&R studies are a powerful tool to identify adaptive mutations from standing genetic variation and thereby provide an excellent means for analyzing the trajectories of selected alleles in evolving populations.

## Results

To minimize the parameter space for the forward simulations we picked commonly used default conditions and varied only the parameter of interest within these defaults. By default we used a base population consisting of 1,000 homozygous genomes that capture the pattern of natural variation found in a population of *D. melanogaster* from Vienna [14], the recombination rate of *D.melanogaster* [15], 3 biological replicates and 60 generations of selection (see Materials and Methods). For every experimental design 10 independent simulations were performed. We simulated 150 codominant ($h=0.5$) loci with two different selection coefficients of $s=0.1$ and $s=0.025$ (see Materials and Methods), which roughly cover the range of selection coefficients detected in experimental evolution studies [16, 17]. Beneficial mutations were randomly picked from the segregating sites in the base population with no distinction being made between the ancestral or derived allele. Simulations were performed for the major autosomes and chromosomal regions with low recombination rates being excluded because they inflate the false positive rate (supplementary results 1.2; supplementary fig. 2). Selected loci were identified by contrasting allele counts between the base and the evolved populations. While several different test statistics have been used in E&R studies (Cochran-Mantel-Haenszel (CMH) test [4], diffStat [6], association statistic [18] and the $F_{ST}$ [8]), we used the CMH test since it showed the best performance (supplementary results 1.1; supplementary fig. 1). The performance of different experimental designs was compared using Receiver-Operating Characteristic (ROC) plots [19] in combination with the partial area under the curve (pAUC) statistic which is a is subregion of the area under the ROC curve [20, 21]. As we are here mostly interested in the power to identify beneficial loci at low false positive rates (FPR) we use the pAUC with a FPR range from 0 to 0.01 [$pAUC = \int_{0}^{0.01} ROC(f)df$].

## Influence of the number of generations

The duration of an experimental evolution study is a key parameter, since it determines the feasibility of an experiment. We varied the number of generations from 10 to 300 which roughly corresponds to experimental lengths from 4 months to 10 years, respectively. As expected, for both selection coefficients, the number of true positives increased with the duration of the selection experiment (Kruskal-Wallis rank sum test with pAUC; $s$=0.1: $\chi^2$=46.07, $p$=5.8$e$−07; $s$=0.025: $\chi^2$=94.72, $p$<2.2$e$−16). Overall, the effect was more pronounced for weakly selected loci than for strongly selected ones (fig. 1; supplementary fig. 8). With fewer generations the differences between strong and weak selection became more pronounced. Based on a $FPR$<0.01 up to 60% of the selected loci can be identified, independent of the selection coefficient. Notably, only a moderate number of generations was necessary to identify a large number of selected sites. For example, 60 generations of selection resulted already in the identification of 48.4% ($s$=0.1) or 36.2% ($s$=0.025) of the beneficial loci.

## Influence of the number of replicates

One of the most powerful features of experimental evolution is the possibility of replication. We evaluated different numbers of replicates ranging from 1 to 20. As expected, increasing the number of replicates had a significant effect for both selection coefficients (Kruskal-Wallis rank sum test with pAUC; $s$=0.1: $\chi^2$=25.45, $p$=0.00011; $s$=0.025: $\chi^2$=54.22, $p$=1.88$e$−10). Similar to the number of generations, also the number of replicates had a more pronounced effect for weakly selected sites (fig. 2; supplementary fig. 9). Interestingly, given a sufficiently large number of replicates, weakly selected loci can be more readily identified than strongly selected ones (Wilcox rank sum test with pAUC[0≤$FPR$≤0.001]; $W$=100; $p$=1.1$e$−05; fig. 2). We reason that this is because neutral variants linked to weakly selected loci have more time to recombine onto neutral haplotypes as opposed to variants on strongly selected haplotypes, which will rapidly become fixed (supplementary results 1.4). While 5 replicates seem to be sufficient for the identification of strongly selected loci, reliable detection of weakly selected ones requires more replicates (fig. 2).

## Influence of the population size

We also modified the number of individuals evolving in each replicate, ranging from 250 to 8,000. For both selection coefficients, we noticed that a larger population size improves the power to identify beneficial loci (Kruskal-Wallis rank sum test with pAUC; $s$=0.1: $\chi^2$=49.80, $p$=1.5$e$−09; $s$=0.025: $\chi^2$=55.61, $p$=9.8$e$−11). Similar to the results for the number of replicates and generations, weakly selected sites benefitted more from an increase in population size (fig. 3; supplementary fig. 10). One complication arising from the comparison of different population sizes is that larger populations contained more starting variation (ranging from 2,011,991 to 3,129,057 SNPs in our study). Thus for a given FPR cutoff the number of significant loci may depend on the population size. To account for this, we repeated the analysis using absolute numbers of significant loci but obtained similar results (supplementary fig. 11).

## Number of chromosomes in the base population

In E&R studies the base population is typically established from isofemale lines, where the number of lines is frequently smaller than the population size. In such cases, multiple females from each isofemale line are used to generate the base population. We tested the influence of the number of chromosomes in the base population by varying the number of chromosomes from 50 to 2,000. The respective base populations contained between 1,482,521 and 2,685,539 segregating loci. The number of chromosomes had a significant influence on the power to identify beneficial loci for both selection coefficients (Kruskal-Wallis rank sum test with pAUC; $s=0.1$: $\chi^2=45.54$, $p=1.1e-08$; $s=0.025$: $\chi^2=39.92$, $p=1.55e-07$). Contrary to our previous results, we found that strongly selected loci benefit more from an increase in chromosome number than weakly selected ones (fig. 4; supplementary fig. 14). The same trend was seen if the absolute number of significant loci is used instead of the FPR (supplementary fig. 13). As neutral variants linked to strongly selected loci have less opportunity to recombine onto neutral haplotypes before becoming fixed (or at least reaching high frequencies) than variants linked to weakly selected loci, the identification of strongly selected loci benefits more from lower initial levels of linkage disequilibrium in the base population (supplementary results 1.5).

## Generating an outbred population prior to the experiment

A frequently used strategy is to generate an outbred population from a limited number of isofemale lines prior to the selection experiment. By propagating the flies for several generations at a large population size, linkage will be broken up, resembling a natural outbred population [6, 22]. We evaluated the efficiency of this procedure by using a population of 1,000 individuals founded by 50 inbred isofemale lines. The populations evolved under neutrality for up to 100 generations. During this neutral evolution phase variation was lost by genetic drift, resulting in a loss of up to 10.7% of the beneficial alleles. We found that neutral evolution prior to E&R has a significant influence on the ability to identify beneficial loci of both selection coefficients (Kruskal-Wallis rank sum test with pAUC; $s=0.1$: $\chi^2=17.02$, $p=0.009$; $s=0.025$: $\chi^2=22.43$, $p=0.001$). For strongly selected loci, the power to identify beneficial loci increased with the duration of the preceding neutral evolution (fig. 5; supplementary fig. 14). For weakly selected loci we noticed a tradeoff and only an intermediate duration of the neutral evolution phase improved the power of the experiment. Nevertheless, in all cases, the improvement in power was modest relative to the other factors evaluated in this study.

## Tradeoff between population size and number of replicates

Because the power to identify beneficial loci was positively correlated with both the population size and the number of replicates, we were interested to determine the optimal strategy given finite resources. For simplicity, we assumed that costs scale linearly with the total number of individuals maintained during the experiment. We analyzed a target of 8,000 flies per experiment, ranging from a single replicate with 8,000 flies to 16 replicates with 500 flies. We found that increasing the number of replicates improved the power to identify strongly selected loci but no effect was observed for weakly selected sites (Kruskal-Wallis rank sum test with pAUC; $s=0.1$: $\chi^2=27.52$, $p=1.6e-05$; $s=0.025$: $\chi^2=0.3$, $p=1.0$; fig. 6; supplementary fig. 15; for $s=0.005$ see supplementary fig. 16). At small population sizes genetic drift has a strong influence on the trajectories of selected loci. This provides an advantage for the identification of strongly selected loci as it may delay rapid rises in frequency of the selected loci and thus provide more opportunities for linked variants to recombine onto neutral haplotypes (supplementary results 1.6).

## Detection limit of beneficial loci with different experimental designs

One essential question is how the experimental design affects the minimum detectable effect size of beneficial loci. We addressed this question by evaluating two extreme experimental designs, one low budget and one high budget design. The low budget design encompasses three replicates, each with a population size of 500, maintained for 60 generations. The high budget design is based on 10 replicates of 2,000 individuals propagated for 120 generations. For both experimental designs we evaluated the power to identify beneficial loci for selection coefficients ranging from 0.5 to 0.001. As expected the selection coefficient had a significant influence on the ability to identify beneficial loci for both experimental designs (fig. 7; supplementary fig. 17; Kruskal-Wallis rank sum test with pAUC; low budget: $\chi^2$=53.15, $p$=3.14$e$−10; high budget: $\chi^2$=55.58, $p$=9.9$e$−11). Very weakly selected loci ($s$=0.001) could not be detected in either experimental design. For all other selection coefficients the high budget design drastically outperformed the low budget design (Wilcoxon rank sum test with *pAUC*; $s$=0.5−0.005: $W$=0, $p$=1.1$e$−05; $s$=0.001: $W$=25, $p$=0.06). Interestingly, for both experimental designs intermediate selection coefficients ($s$=0.05) performed best (fig. 7; supplementary fig. 17).

Another way to evaluate the power of different designs is the absolute number of selected loci found among the most significant SNPs. The results for the high budget design were particularly encouraging since among the 10 most significant loci up to 7 true positives were detected (Table 1). Additionally, of the 100 most significant loci about 50 were true targets of selection. On the other hand, the low budget design only resulted in moderately encouraging results. When the 10 most significant loci were considered at most 2 true positives were found. Considering a larger number of most significant SNPs only improved the result marginally (Table 1). To allow for a more intuitive comparisons of the two experimental designs with actual experimental evolution studies we have visualized the results using Manhattan plots (low budget design: supplementary figures 18,19,20,21,22,23; high budget design: supplementary figures 24,25,26,27,28,29).

## Discussion

Our simulations revealed an interesting difference between strongly and weakly selected sites. Experimental designs with low power typically favor strongly selected sites. This advantage is, however, frequently lost with more powerful designs. The power of an experimental design ultimately depends on the signal-to-noise ratio (the signal is the average log-transformed p-value of selected sites and the noise the average log-transformed p-value of non-selected sites). Strongly selected loci were characterized by a strong signal that was easily recognized even with less powerful experimental designs. In contrast the signal of weakly selected loci needs to be detected with powerful experimental designs and distinguished from the background noise caused by neutrally evolving loci. Interestingly, once weakly selected loci are detected the noise caused by hitchhiking variants is lower than for strongly selected loci. This different behavior of strongly and weakly selected loci is further illustrated in the online supplement by analyzing signal and noise separately (supplementary results 1.4).

We showed that the following key factors could increase the sensitivity of an experimental design: 1) larger population size, 2) more replicates, 3) increasing number of generations, and 4) chromosome diversity at the beginning of the experiment. The benefits of higher chromosome diversity largely results from uncoupling linkage between selected and neutral loci. A longer duration of the selection experiment and a higher number of replicates, however, have a two-fold influence. On one hand they increase the sensitivity, particularly for weakly selected loci, by making consistent allele frequency change across replicates more apparent. On the other hand they will increase the number of recombination events that occur during the experiment which will also reduce the noise by uncoupling selected loci from neutral linked ones. A larger population size has a similar effect: by reducing genetic drift even weaker signals of selection can be detected. Furthermore, the number of recombinant haplotypes is increased, uncoupling true- and false-positives.

Selected loci were detected with the CMH-test, which is based on contrasting allele frequency differences between the base population and the evolved populations. While our results are robust compared to similar test statistics (supplementary results 1.7), those based on time series data may lead to different conclusions. Time series data can be easily obtained for E&R studies by sequencing the evolving populations at certain intervals, which could improve the power to identify beneficial loci (e.g.: [23]). Currently, suitable test statistics for time series data from E&R studies are still lacking, but once such tests are available it will be important to re-evaluate whether the recommendations derived here are still valid. Furthermore, these datasets will require evaluation of additional factors such such as the optimal number of sampled time points and the optimal interval between.

In this study, we have identified beneficial loci based on the actual allele frequencies in the populations. In an E&R study exact allele frequencies require sequencing of all individuals in a population separately. While technically feasible, budget constraints preclude this approach. Rather, E&R studies have relied on Pool-Seq as a more cost effective approach [24], to estimate allele frequencies [4-8, 18]. Since coverage is the key parameter determining the ability to detect allele frequency differences, the well-described phenomenon of coverage heterogeneity along chromosomes [25, 26] provides an additional challenge for the analysis of E&R studies: selected sites in genomic regions with a lower coverage will be less likely to be detected than loci with a higher coverage. Our results are, however, robust with respect to the sequencing technology and are not affected by problems that are common with the NGS technology like sequencing errors or uneven coverages.

To reduce the parameter space we performed all simulations with 150 beneficial loci, as we found this to be the highest number of loci that could be simultaneously selected without disproportionally reducing the efficacy of selection (supplementary results 1.3; supplementary fig. 3). Fewer beneficial loci will slightly increase the efficacy of selection of strongly selected loci, but there seems to be little effect for weakly selected loci (supplementary fig. 3). The power to identify beneficial loci in an E&R study with few strongly selected loci may therefore be slightly higher than suggested by our results.

E&R is a powerful tool to pick up even weakly selected loci, but the challenge of E&R studies lies in the distinction between true targets of selection and false positives. Our simulation results based on a small number of replicates and moderate populations sizes mirror those from previously published experimental studies, which also identified a large number of candidate loci [4-6, 8]. Depending on the strength of selection and the experimental design, neutral loci linked to the target of selection could create high levels of false positives (supplementary fig. 24). On the other hand it is also possible that false positives are not flanking targets of selection (supplementary fig. 28). Hence, it is essential for E&R studies to use an experimental setup which assures the maximum enrichment of truly selected sites among the top candidates. Since the true number of selected loci is not known in experimental studies our simulation results are an important guideline for future E&R studies. Based on our computer simulations, we recommend to maximize the number of replicates, the population size, the duration of the experiment and the number of chromosomes seggregating in the base population. We advise to prioritizing replication over population size for species such as *Drosophila*, where maintenance of large population sizes is resource intensive. Furthermore, we note that replication reduces the damage if one population is lost, contaminated or otherwise compromised during the experiment.

In summary, we showed that selected loci with a selective advantage as low as *s*=0.005 can be reliably identified at the nucleotide level when using an optimal experimental design. Even when a large number of loci are selected simultaneously, up to 56% can be detected in the laboratory with a low rate of false positives. Our computer simulations suggest that, provided an adequate experimental design, E&R studies are a powerful tool to identify adaptive mutations from standing genetic variation and thereby will be a major contributor in elucidating the dynamics of adaptation from standing genetic variation.

## Materials and Methods

## Simulation software

We developed MimicrEE (Mimicing Experimental Evolution), a forward simulation tool designed to model experimental evolution for fully sequenced genomes in the presence of multiple selected loci. This software comprises several Java applications that utilize the same core library (mimcore). MimicrEE performs forward simulations for a population of diploid individuals. These diploid genomes are provided as haplotypes with two haplotypes constituting a diploid genome (further details are given in the manual: https://code.google.com/p/mimicree/wiki/Manual). MimicrEE can handle large populations with several million segregating sites. So far it has successfully been tested with 8,000 diploid individuals each having 4 million polymorphic loci. Memory requirements scale linearly with the population size and number of loci. During the forward simulations the population size is kept constant. A list of selected loci may be provided and in case that no selected locus is specified neutral simulations will be performed. For each selected locus the selection coefficient (*s*), the heterozygous effect (*h*), and the nucleotide of the non-selected allele needs to be provided ($w_{11}$).

The fitness of the heterozygous and homozygous individuals is given by: $w_{11}=1$, $w_{12}=1+hs$, $w_{22}=1+s$ (see also [28]); We assume an additive model when multiple selected loci are specified.

No *de novo* mutations are considered as we are specifically interested in adaptation from standing genetic variation. The forward simulations are performed with non-overlapping generations of hermaphrodites with selfing being excluded. At each generation matings are performed, where mating success (i.e.: number of offspring) scales linearly with fitness, until the total number of offspring in the population equals the targeted population size (fecundity selection). Each parent contributes a single gamete to the offspring wherein crossing over events are introduced according to the specified recombination rate. The recombination rate can be provided for arbitrarily sized windows. MimicrEE provides either haplotypes (MimicrEEHaplotype) or a summary of the allele frequencies (MimicrEESummary) at generations specified by the user as output.

# Forward simulations

We simulated 8,000 chromosomes with fastsimcoal v1.1.8 [29], that capture the pattern of natural variation of a *D. melanogaster* population. The diversity ($\Theta_\pi$) of a *D. melanogaster* population, captured in Vienna in fall 2010 [14], was estimated with PoPoolation v1.2.1 [30]. The recombination rate was obtained from the *D. melanogaster* recombination rate calculator v2.2 [15]. Recombination rate and diversity were specified in 100kb windows, where the diversity was provided as mutation rate parameter adjusted to the pairwise distance in a window for a population size of 750,000 ($\mu=\Theta/4N_e$). We excluded the X-chromosome and low recombining regions (<1*cM/Mb*), including the entire $4^{th}$ chromosome, from the analysis (supplementary results 1.2; supplementary fig. 2) Thus we performed our simulations only with the high recombining regions of the chromosomes 2L, 2R, 3L, and 3R; To build a population for the start of the experimental evolution simulation (base population) we randomly sampled chromosomes (default=1,000) from the 8,000 simulated chromosomes. If the number of chromosomes sampled for the base population was smaller than the targeted population size, homozygous individuals were formed to mimic an inbred isofemale line. We found that up to 150 beneficial loci could be included in a simulation run without disproportionally reducing the efficacy of selection (supplementary results 1.3; supplementary fig. 3). Therefore, we performed all simulations with 150 randomly drawn beneficial loci, with no distinction being made between the ancestral or the derived allele. To increase the probability of detecting the selected allele, the starting frequency of the selected loci was not allowed to exceed 80%. All selected loci were codominant (*h*=0.5) and, if not mentioned otherwise, had a selection coefficient of *s*=0.1 or *s*=0.025. The recombination rate for the forward simulations was also obtained from the *D. melanogaster* recombination rate calculator v2.2 [15]. Since males in *Drosophila* are not recombining and we simulated hermaphrodites we divided the female recombination rate by 2. Forward simulations were performed with MimicrEESummary for the given numbers of generations (default=60). Biological replicates for every experiment were acquired by repeating forward simulations several times (default=3). Confidence intervals for all experimental designs are based on 10 independent repetitions using the same parameters, but different selected loci. Note that each repetition included biological replicates.

# Statistical analysis

We used the the Cochran-Mantel-Haenszel test (CMH-test; [31]), implemented in PoPoolation2 r185 [32], to identify selected loci as this test showed the best performance (supplementary results 1.1; supplementary fig. 1). The CMH-test is based on a meta-analysis of contingency tables, where one contingency table is used for each biological replicate. These contingency tables contain the allele counts of the base and the evolved populations. We wrote custom scripts in Python for calculating the diffStat [6], the association statistic [18] and the average pairwise $F_{ST}$. For statistical analysis we used the R programming language [33] and the library ROCR [34].

## Availability

MimicrEE has been released under the Mozilla Public License 1.1 and is available from https://code.google.com/p/mimicree/

## Author's contributions

CS and RK conceived the study. RK wrote the software and conducted the analysis. RK and CS wrote the paper

## Acknowledgements

We thank all members of the Institute of Population Genetics for feedback and support. We especially thank Liang Leng, Andrea Betancourt, Susanne Franssen and Raymond Tobler for discussions and critical comments. This work was supported by the ERC grant "Archadapt" and Austrian Science Funds (FWF) grant P22725.

# Figures

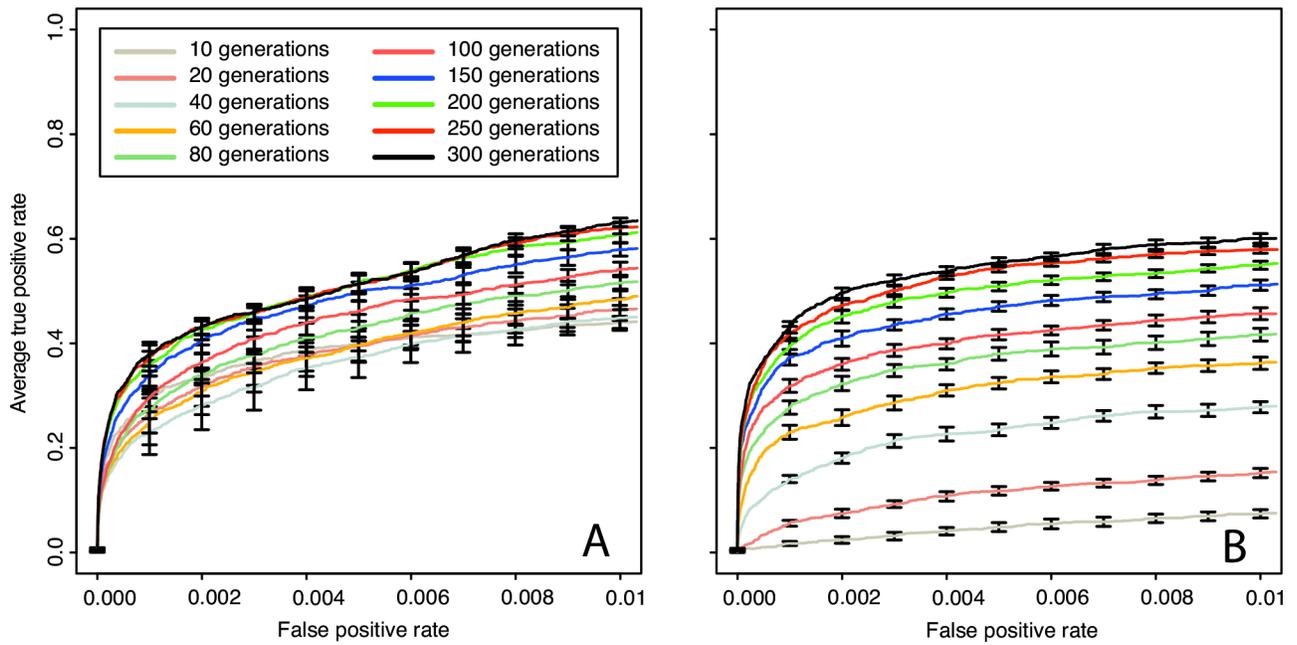

Figure 1: Influence of the number of generations on identification of beneficial loci with selection coefficients of $s$=0.1 (A) and $s$=0.025 (B)

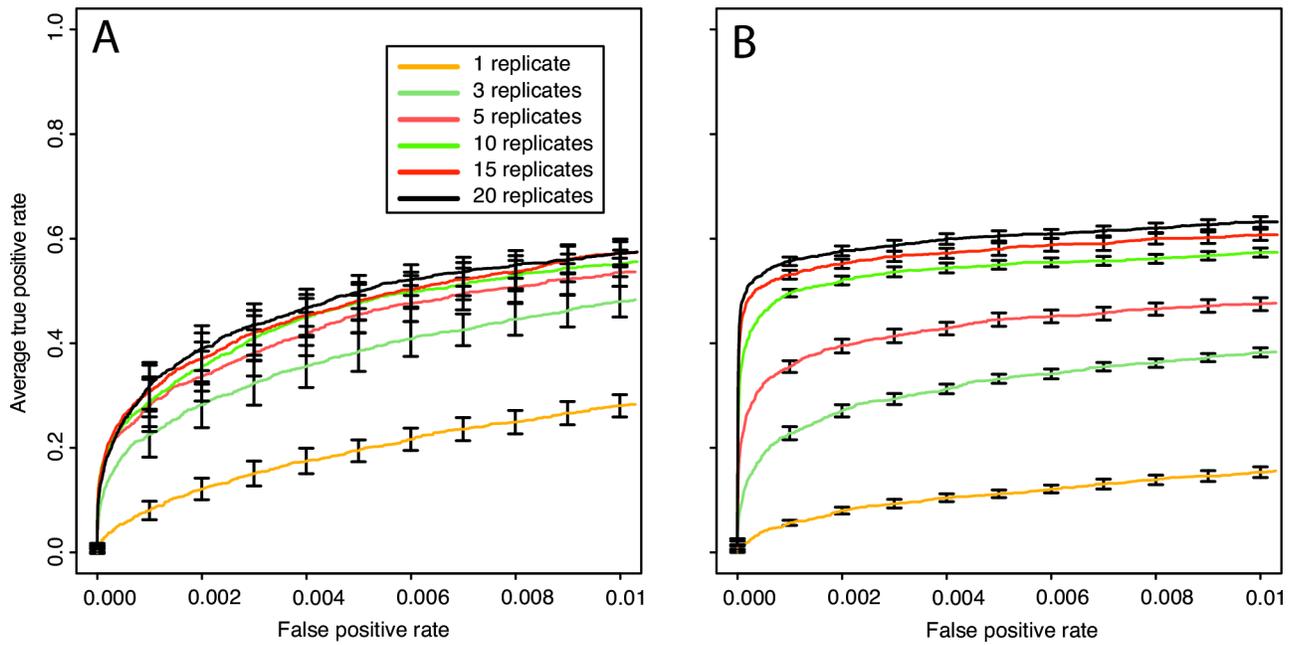

Figure 2: Influence of the number of biological replicates on identification of beneficial loci with selection coefficients of *s*=0.1 (A) and *s*=0.025 (B)

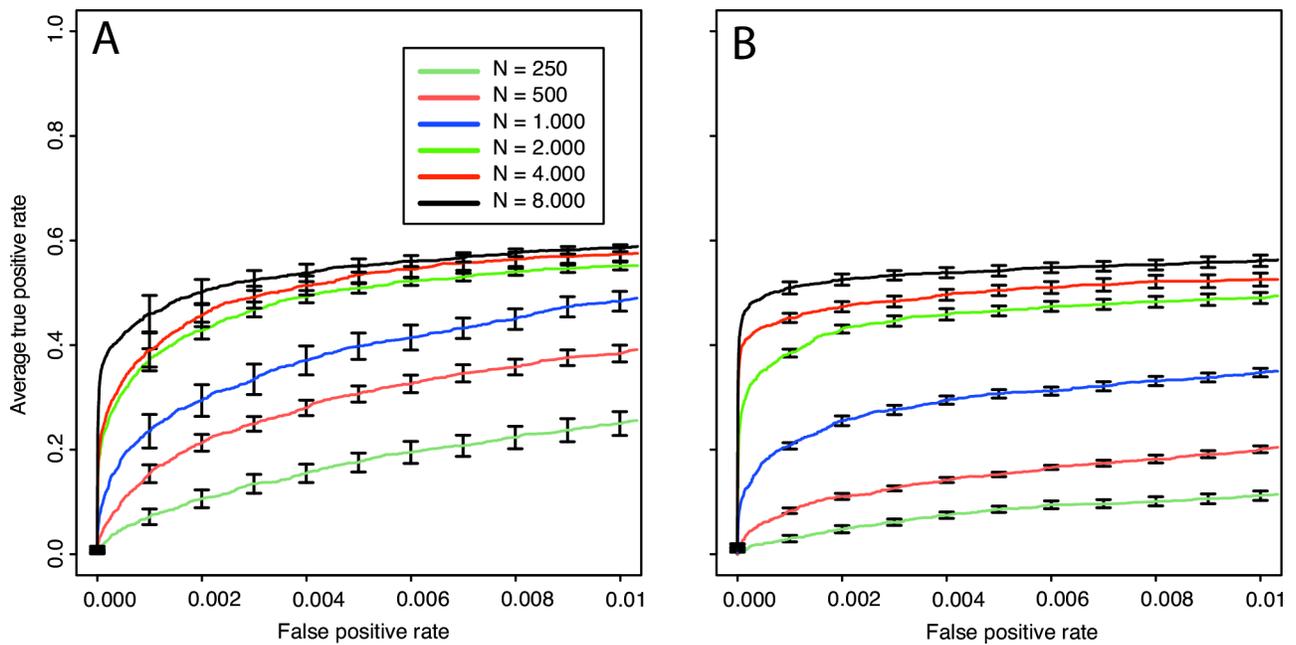

Figure 3: Influence of the population size on identification of beneficial loci with selection coefficients of *s*=0.1 (A) and *s*=0.025 (B)

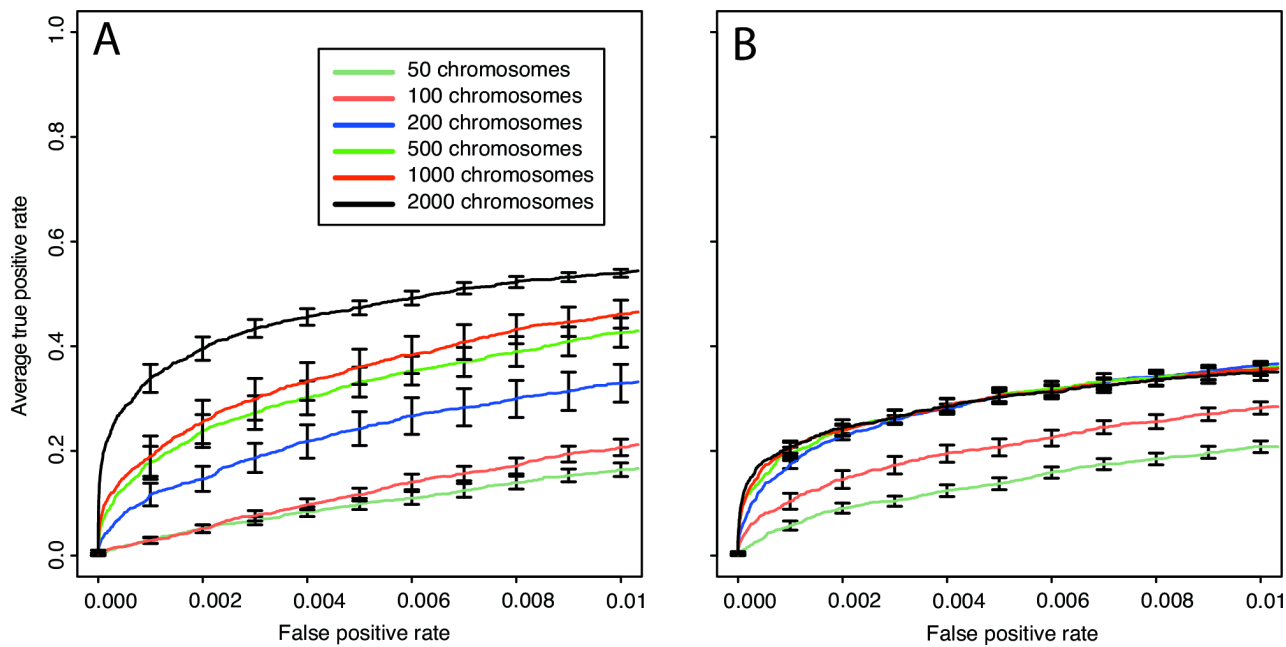

Figure 4: Influence of the number of starting chromosomes in a population with size of 1,000 on the identification of beneficial loci with selection coefficients of $s$=0.1 (A) and $s$=0.025 (B); Only homozygous individuals were used except in the population with 2,000 chromosomes

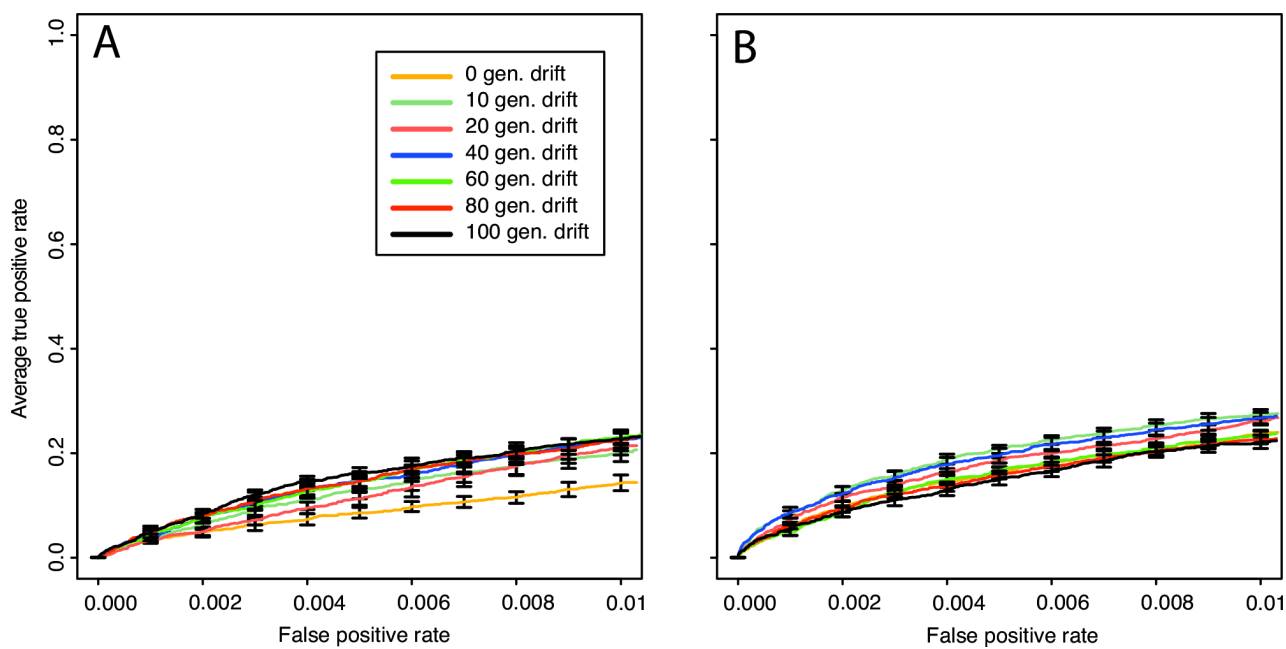

Figure 5: Influence of drift prior to experimental evolution on the identification of beneficial loci with selection coefficients of *s*=0.1 (A) and *s*=0.025 (B); A chromosome number of 50 and a population size of 1,000 was used. In the base population every chromosome was thus present in 20 homozygous individuals.

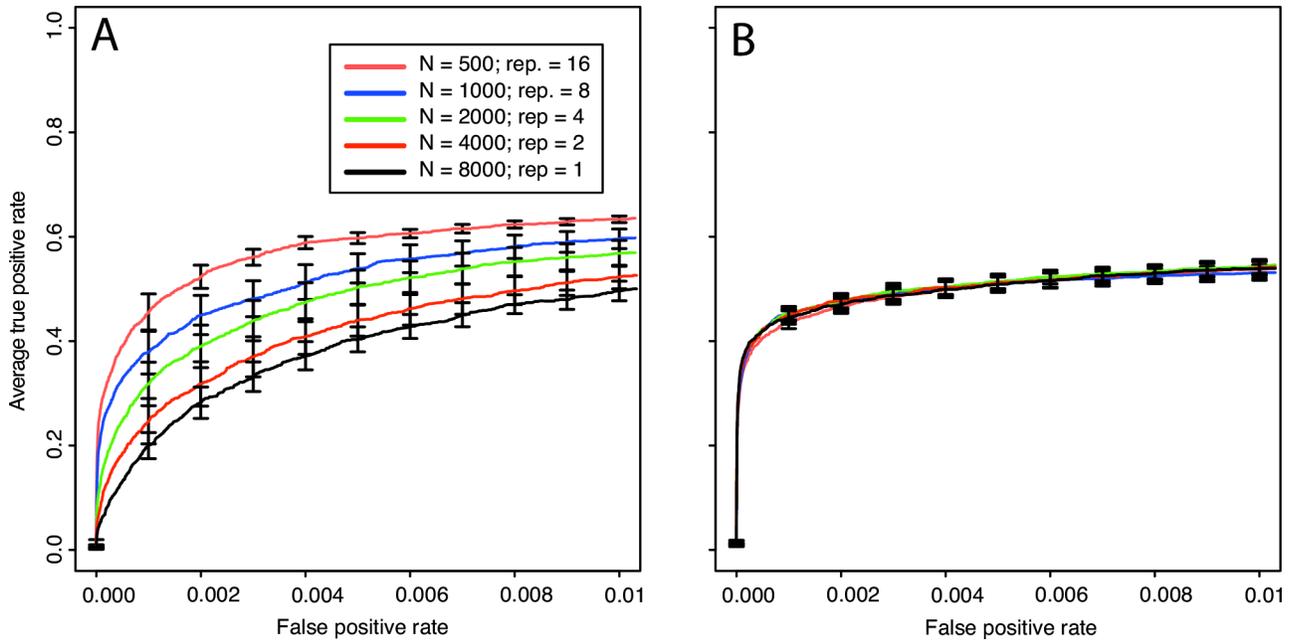

Figure 6: Influence of the tradeoff between population size and number of replicates on the identification of beneficial loci with selection coefficients of *s*=0.1 (A) and *s*=0.025 (B)

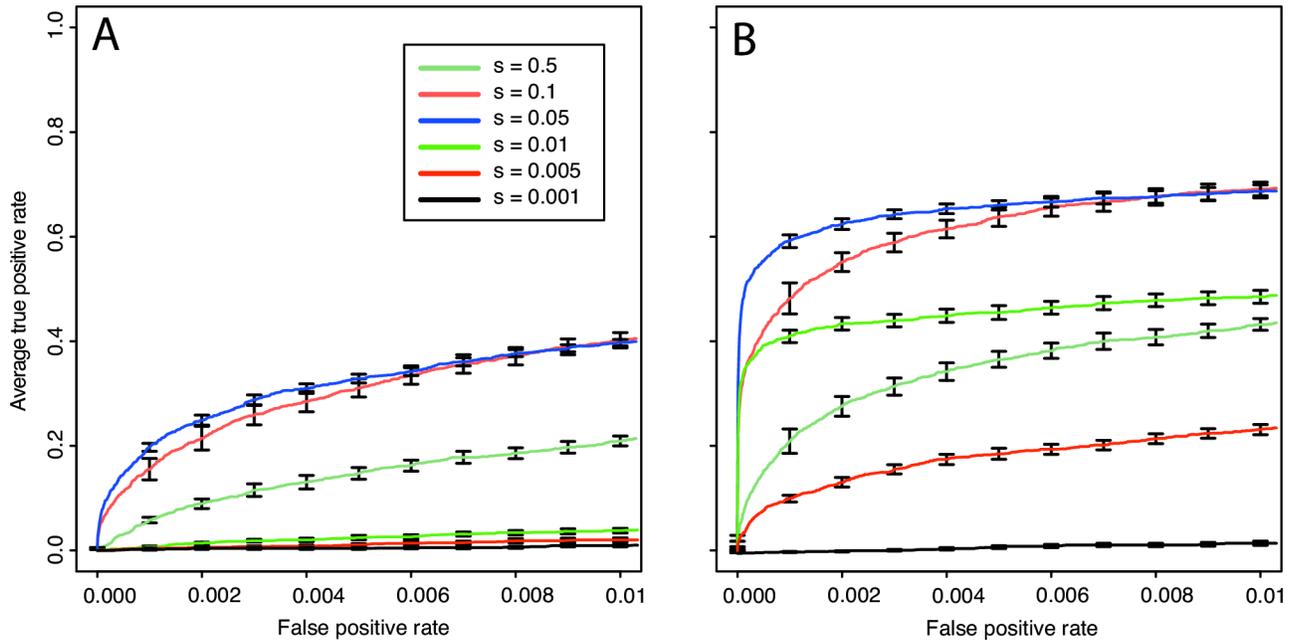

Figure 7: Power to identify beneficial loci of different effect sizes with a low budget (A) and a high budget (B) study design; See text for definition of low budget and high budget study design.

# Tables

Table 1: Power to identify beneficial loci of different selection coefficients (s) with a low budget (lb) and a high budget (hb) study design. Reported values are average numbers of true positive identified with the given number of most significant SNPs. See text for definition of low and high budget study design.

| s | | 10 | 20 | 50 | 100 | 200 | 500 | 1,000 | 2,000 | 5,000 | 10,000 |
|---|---|---|---|---|---|---|---|---|---|---|---|
| 0.5 | lb | 0.3 | 0.5 | 0.8 | 1.7 | 2.8 | 5.4 | 9 | 13.5 | 22.5 | 29.8 |
| | hb | 1.5 | 2.6 | 4.7 | 7 | 9.5 | 15.5 | 21.6 | 29.6 | 42.9 | 53 |
| 0.1 | lb | 2.4 | 3.4 | 4.8 | 6.5 | 8.4 | 12.4 | 16.9 | 23.3 | 35.9 | 46 |
| | hb | 6.6 | 13.1 | 27.8 | 37.8 | 44.6 | 53.3 | 59.6 | 67.8 | 82.2 | 91.9 |
| 0.05 | lb | 2.4 | 3.5 | 5.5 | 8.2 | 11.7 | 16.6 | 21.7 | 29.4 | 39.7 | 49.1 |
| | hb | 7.7 | 15.4 | 35.3 | 52.4 | 65 | 76.4 | 80.9 | 86.4 | 93.2 | 97.7 |
| 0.01 | lb | 0 | 0 | 0.1 | 0.1 | 0.1 | 0.3 | 0.5 | 1 | 2.5 | 3.5 |
| | hb | 7.4 | 14.6 | 30.6 | 40.5 | 48 | 53.6 | 56.9 | 60.9 | 65.2 | 67.6 |
| 0.005 | lb | 0 | 0 | 0 | 0 | 0 | 0.1 | 0.4 | 0.5 | 0.9 | 1.5 |
| | hb | 1.7 | 2.2 | 3.2 | 4 | 5.2 | 7.7 | 11.1 | 14.4 | 20.1 | 26.8 |

| 0.001 | lb | 0 | 0 | 0 | 0 | 0 | 0 | 0.2 | 0.3 | 0.6 | 0.6 |
| | hb | 0 | 0 | 0 | 0 | 0 | 0.1 | 0.2 | 0.3 | 0.6 | 1.3 |

## Supplementary Files

Supplementary files are available at http://drrobertkofler.wikispaces.com/GuidelinesSupplement

## Supplementary file 1

A pdf file containing supplementary text and the supplementary figures.

## Supplementary file 2

An excel worksheet containing all *AUC* and *pAUC* values used in this work